\documentclass[18pt,aps,nofootinbib]{revtex4}
\usepackage{amsmath}
\usepackage{graphicx}
\usepackage{slashed}
\usepackage{color}
\usepackage[normalem]{ulem}

\def\gsim{\gtrsim}

\begin{document}
\title{Proton fraction in neutron star matter: Dynamical mean-field approach.}
\author{Maxim Velikanov$^{1}$, Alexey N. Rubtsov$^{1}$, Boris Krippa$^{2}$}
\affiliation{$^1$ Russian Quantum Center, Moscow, Russia}
\affiliation{$^2$Department of Physics, Imperial College London, London, UK}
\date{\today}
\begin{abstract}
Dynamical mean field theory (DMFT) is used to study neutron matter, both with and without admixture of the proton fraction. The system is approximated by the lattice Habbard model. The corresponding equation of state as a function of temperature/density/asymmetry is investigated. The results are compared with the standard mean field (MF) approach where the effect of local correlations is neglected. Whereas the influence of the correlations on the properties of a pure neutron matter is found to be moderate, it becomes strong when the proton admixture is taken into account. In particular, we calculate the proton fraction, energy density and pressure in outer core of neutron stars, taking into account the beta equilibrium condition. The DMFT predicts that the proton fraction is several times the MF based calculations, whereas the DMFT results for energy density and pressure are 30-40\% lower then the corresponding MF estimates. Physical implications of our findings for a neutron star dynamics are discussed.

%In particular, our predictions for the proton fraction and for the pressure of the $\beta$-equilibrated neutron matter of the outer part of the core of the neutron star are respectively one order of magnitude higher and 30-40\% lower than the results of the standard mean field consideration.%
%taking into account  the local correlations is 
%quite significant for the resulting equation of state. The implications for the neutron stars are discussed.
\end{abstract}
\maketitle
\begin{section}
{\bf INTRODUCTION}
\end{section}
 The study of strongly correlated nuclear many-body systems is one of the key problems of the strong interaction physics described by QCD.  At low energy  QCD can be reduced to an effective theory containing  the effective degrees of freedom instead of the fundamental ones  \cite{EFT}. An  extension of such theory to the  finite density would allow a description of  a wide range of phenomenon occurring in nuclear matter, heavy-ion collisions and in the neutron stars interior \cite{Lat}. 

The most straightforward way of calculating observables in QCD from the first principles is to use the standard Monte-Carlo methods for discretisation and numerical integration of the corresponding path integral. It allowed to calculate many observables both in vacuum and at finite temperature.

However, at finite density a well known problem has been an inability to perform lattice  ab initio simulations  using Monte Carlo approaches since the fermion determinant becomes complex in the presence of finite baryon chemical potential, the notorious “sign problem.” In spite of the significant progress which has been made in recent years in overcoming some of these problems, most of the numerical studies in this area focus on small chemical potential, analytical techniques, and investigation of model systems. However, away from the region along the temperature axis, the QCD phase diagram remains, for the most part, inaccessible for the direct methods.

One way to tackle this problem is to use effective models, adjusted to take into account finite density corrections. However, most of them are treated using mean-field approximation where the effects of correlations are neglected. Another problem the effective models are facing is evaluation and  inclusion of the beyond-mean-field corrections.

A possible alternative to a direct numerical simulation and effective models is to consider  three-dimensional lattice models with local interaction, similar to those adapted in condensed matter physics like the attractive Hubbard model.

In the current work we will apply the Habbard model to study strongly correlated nuclear many-body systems at zero temperature. The low energy nucleon-nucleon (NN) interactions are characterized by scattering length $a_S$ which is by an order of magnitude larger than the effective radius $r_e$ so that the neutron/nuclear matter is close to unitary regime where the Fermi momentum remains the only length scale and the interaction can be viewed as being of zero range. Clearly the  neutron/nuclear matter is not exactly at the unitary point so that a finiteness of the interaction range plays a certain role but choosing the model with the zero range interaction seems to be a reasonable starting point \cite{Gries} at least at not too large densities. 

  We use the dynamical mean field theory (DMFT)  \cite{Geo} to treat Habbard model for neutron and asymmetric nuclear matter. DMFT is the generalisation of the standard mean field theory where a lattice problem is reduced into a self-consistent impurity problem. In this approach the impurity is influenced by a bath with which the impurity hybridizes. The main approximation made in the standard DMFT is to treat the lattice self-energy as  a momentum-independent (local) quantity. This approximation becomes exact in the limit of lattices of infinite dimensionality \cite{Metz}. DMFT has previously been used in the context of the lattice models for the solid state physics \cite{Geo} and dilute Fermi gas \cite{Bar} and has proven to be a very successful in describing various experimental data. We will obtain and analyse the corresponding equation of state and apply the results to the physics of neutron stars. Regardless of the interaction strength, the zero temperature many-fermion system with attractive interaction exhibits an instability which leads to the rearrangement of the ground state of the system and we allow for a corresponding symmetry breaking in our DMFT consideration.

In the case of neutron stars, it is worth to consider the admixture of the proton fraction, which while being small nevertheless plays an important role in the underlying dynamics \cite{Pek}. It influences the cooling rate of the neutron stars which is defined by the URCA process \cite{URCA}. There is a minimum proton fraction for which the URCA process can occur $\simeq 12\%$. A reliable estimate of this fracture is still in open question in neutron star physics. A determination of this parameter is one of the main purposes of this paper. Also it defines the density dependence of the symmetry energy which in turn is crucial for the existence of some exotic phases like pasta one in the density region where the stellar crust turns into outer part of the core.

The paper is organized as follows. First we will introduce the attractive Hubbard model and describe the Monte-Carlo  approach used to solve the corresponding impurity problem. The next section is devoted to implementation of DMFT for the neutron/nuclear matter. Then we  present the results of both static and dynamical mean-field approaches and compare them. Both symmetric and asymmetric nuclear matter are considered. We summarize and describe future plans in the last section.
\vskip0.5cm
\begin{section}
{\bf LATTICE HAMILTONIAN}
\end{section}

The central step in our approach to simulating neutron/asymmetric nuclear matter is to map it to the lattice. Then we apply Dynamical Mean Field Theory (DMFT) to the resulting lattice model, which means that we are supposed to use the Hubbard model with local on-site interaction thus having just a few adjustable parameters - the lattice geometry, hoping parameters, and components of the interaction strength tensor $U$. The  Hamiltonian of the Hubbard model with neutrons and protons is
\begin{equation}
\label{HubbardM1}
H = \sum \limits_{j,j', l, s} (t_{jj'} - \mu_l \delta_{jj'}) c_{j l s}^\dagger c_{j' l s} + \sum\limits_j H_{int}[c^\dag_j, c_j],
\end{equation}
where $t_{ij}$ describes hoping between different sites $j, j'$, indices $l=n,p$ and $s=\uparrow, \downarrow$ number isospin and spin projections, $c^{\dag}, c$ are creation-annihilation operators, $\mu_l$ are chemical potentials, and $\delta_{jj'}$ is the Kronecker's symbol.  $H_{int}$ is the local Hubbard interaction, which obeys the following structure determined by the symmetry: 
\begin{eqnarray}
H_{int}[c^\dag_j, c_j] = U_{nn} (c_{n\uparrow,j}^\dagger c_{n\downarrow,j}^\dagger c_{n\uparrow,j}c_{n\downarrow,j}) + U_{pp} (c_{p\uparrow,j}^\dagger c_{p\downarrow,j}^\dagger c_{p\uparrow,j}c_{p\downarrow,j})  + U^s_{np}\frac{1}{2} (c_{n\uparrow,j}^\dagger c_{p\downarrow,j}^\dagger  - c_{n\downarrow,j}^\dagger c_{p\uparrow,j}^\dagger) (c_{n\uparrow,j} c_{p\downarrow,j}  - c_{n\downarrow,j} c_{p\uparrow,j})+\nonumber \\ + U^t_{np} \left( c_{n\uparrow,j}^\dagger c_{p\uparrow,j}^\dagger c_{n\uparrow,j} c_{p\uparrow,j} + c_{n\downarrow,j}^\dagger c_{p\downarrow,j}^\dagger c_{n\downarrow,j} c_{p\downarrow,j} + \frac{1}{2} (c_{n\uparrow,j}^\dagger c_{p\downarrow,j}^\dagger  - c_{n\downarrow,j}^\dagger c_{p\uparrow,j}^\dagger) (c_{n\uparrow,j} c_{p\downarrow,j}  - c_{n\downarrow,j} c_{p\uparrow,j}) \right)  
\end{eqnarray}
Here $U_{nn}, U_{pp}, U^s_{np}, U^t_{np}$ are coupling constants of the neutron-neutron, proton-proton, neutron-proton singlet and triplet interaction, respectively. Simple solvable regimes of the model are $U\rightarrow  0$, 
when we have a non-interacting system, and $t\to 0$, when the system is in the atomic limit.

The lattice Hubbard model with the local interaction allows us to handle the most important $^1S_0$ pairing mechanism, which dominates for the neutron matter at low density. However, we note that pairing at higher angular momentum, which may occur at higher densities, is not included in the formalism and is therefore beyond the scope of our consideration.

Now let's formulate the DMFT equations. We choose one of the lattice sites and focus on a local physics on this site. The rest of the lattice is replaced by a non-correlated effective bath to be defined in a self-consistent way. The problem of a single site coupled to a quadratic bath is called Anderson impurity model (AIM). It is described by the action
\begin{equation}\label{AIM}
    S^{AIM}[c^\dag_j, c_j]= \sum_{l, s, \omega} c^\dag_{j \omega l s} (- i \omega +\Delta_l(i \omega)) c_{ j \omega l s} + \frac{1}{2} \sum_{l, s, \omega} \tilde \Delta_{\omega l} (c^\dag_{j \omega l s} c^\dag_{j \omega l \bar s} + c_{j \omega l s} c_{j \omega l \bar s}) + S^{int}[c^\dag_j, c_j].
\end{equation}
Here $\Delta_l(i \omega)$ is the hybridization function, describing the effect of the bath. The difference between the usual (``static'') mean field and the DMFT is that the latter fully takes the on-site correlations into account. Consequently, the Gaussian bath carries time-dispersion effects, and $\Delta$ depends on the fermionic Matsubara frequency $i \omega$. Different chemical potentials for neutrons and protons make $\Delta$ isospin-dependent. Possible pairing effects require an account of the abnormal components of the hybridization $\tilde \Delta$ \cite{Abr}. Finally, $S^{int}$ is the part of action corresponding to the on-site interaction $H^{int}$. 

Given the hybridisation function, one numerically obtains the AIM's Green's function and, consequently, the self energy:
\begin{align}
    \label{DMFT1}
    &\Delta(i\omega) \overset{\text{AIM solver}}{\longrightarrow} G_{imp}(i\omega) \\
    \label{DMFT2}
    &\Sigma_{imp}(i\omega) = \left(i\omega + \mu - \Delta(i\omega)\right) - G_{imp}^{-1}(i\omega).
\end{align}
Here the abnormal components should be taken into account, so that $\Delta(i\omega)$, $G^{imp}$, and $\Sigma^{imp}$ in the above formulas are tensors with spin-isospin-Nambu indices. The AIM numerical solver used in our consideration is described in the next section. 

The key assumption of the DMFT method is that local single-particle dynamics of the lattice can be approximated by that of the AIM. Thus the lattice self energy is local and equals to $\Sigma^{imp}$, and the local part of the lattice Green's function equals to $G_{imp}$:
\begin{equation}
        \label{DMFT3}
    G_{imp}(i\omega) = \int \frac{  d^3k}{i \omega - \varepsilon_k + \mu - \Sigma_{imp}(i\omega)},
\end{equation}
where the dispersion law $\varepsilon_k$ is the Fourier transform of the hopping term $t_{jj'}$. Formula (\ref{DMFT3}) is the DMFT self-consistency condition. In practice, the DMFT calculations involve an iterative change of $\Delta, \tilde \Delta$ and calculation of (\ref{DMFT1}, \ref{DMFT2}) performed until the condition (\ref{DMFT3}) is fulfilled.

As we have mentioned, the DMFT equations are well-defined for a lattice system, so that the integration in (\ref{DMFT3} goes over the Brillouin zone, and $k$ is the quasi-momentum. Working with a continuous system requires an introduction of certain lattice discretisation. The most prominent effect of the lattice with the period $l$ is that it introduces the upper cut-off at $k \sim 1/l$. This cut-off should at least be  several times larger than the Fermi momentum, which puts the requirement $l \lesssim \pi/k_F$. However, there is also a condition which forbids using of a very small lattice period. Indeed, the DMFT assumes that correlations are local at lattice sites. Therefore, the lattice period should be chosen similar to or larger than the correlation length of the neutron matter. The later quantity can be roughly estimated from the effective interaction length $r_e$, so that the requirement is $r_e \gsim l \gsim \pi/k_F$. 

Thus, a properly chosen lattice delivers a balance between the approximations in fermion kinetics (smaller $l$ is better) and their correlations (larger $l$ is better). For practically interesting regimes, the value of $r_e k_F$ is about 0.5, which leaves not much space for possible values of $l$ and makes a proper choice of the lattice discretisation crucial. While the volume of the Brillouin zone is effectively restricted by the inverse correlation volume, one should look for a lattice which delivers a dispersion law close to the original parabolic one within as large range of $k$ as possible. 

In our consideration, we follow the Debye prescription: introduce an effective spherical Brillouin zone $k<k_{max}$ with $\varepsilon_k=\frac{k^2}{2 m}$ in all points. Of course this does not correspond to any physical lattice discretisation. However, we have all the parameters entering the DMFT equations. Except for the integral in (\ref{DMFT3}), the only point where properties of the lattice appear is the interaction term of the impurity model. It is proportional to the volume of the elementary cell of the lattice, which should be taken equal $\frac{6 \pi^2}{k_{max}^3}$. The same method was used previously \cite{Pri} and indeed demonstrated better results than physical lattice discretisations.

Finally, the effective interaction $U$ should be determined. Obtaining the coupling constants involves fitting the available experimental data for nucleon-nucleon system. There exists two possible strategies to achieve this. First,  one can use the corresponding low energy experimental data like scattering length $a_s$ and effective radius and then exploit it to extract the values of the coupling constants. This is a standard procedure in the EFT for low-energy nucleon-nucleon interaction \cite{Kap} , which in this case is  just a way to reformulate the effective range expansion.
Being completely justified for nucleon-nucleon interaction in vacuum, where the zero momentum limit can easily be defined
 it is not really suitable for the finite density calculations in the region   $k_F \gtrsim 0.4 \; \text{fm}^{-1}$ which is the region we focus on in this paper.

The other possible option to extract the values of the coupling constants $U_{ij}$ is to use the experimental phase shifts at some fixed relative momentum. This is similar to  the procedure used in ref. \cite{Cla}. 
We have chosen the value
\begin{equation}
\left\langle |\mathbf{k}|\right\rangle_{\{|\mathbf{k}|<k_F\}} = \frac{3}{4} k_F
\end{equation}

The value of the couplings can be extracted from the following expression for the scattering amplitude
\begin{equation}
%\label{matchingoption2}
T =\frac{4\pi}{M} \frac{1}{\left( \cot \delta_{s}(k) + i k\right)},
\end{equation}
where $k$ is some typical low-energy scale (momentum) and $\delta_s$ is the s-wave phase shift. We will use this option for calculations of the equation of state in the realistic neutron matter case. 

For a numerical solution of the impurity problem (\ref{AIM}), we employ a CT-INT Quantum Monte Carlo code \cite{MC}. The algorithm relies on a stochastic walk in the space of Taylor series for the partition function of the impurity problem $Z^{AIM}=\int e^{-S^{AIM}[c^\dag_j, c_j]} D c^\dag_j, c_j$ with respect to $S^{int}$. A probability of each Monte Carlo step was estimated as a ratio of determinants composed of normal and abnormal Green's functions calculated for the non-interacting part of the impurity problem.  Whereas
we followed a generic CT-QMC framework outlined in \cite{MC}, there are several comments regarding the practical performance of the method for the particular system under study. We did not observe a significant sign problem in our calculations, although the presence of spin-flip terms and anomalous averages makes the CT-INT weights not necessarily positive-defined. This can be attributed to an attractive character of the interaction. Additionally we note that for a single-component system with an attraction the sign problem is absent, so that one does not expect that a small proton admixture would result in a severe sign problem.

However, we observed a slow-down problem related to a complex landscape of the weights in the perturbation series for $Z^{AIM}$. We detected that it can be partially cured by splitting the series into sectors with respect to the type of Green's function appearing in the given term. For example, one of the sectors includes all terms composed of neutron-neutron Green's functions only, other includes only ``proton''
propagators, and all possible kinds of the Green's function appear in the the last sector. As one can observe, for the four possible spin-isospin projections there are $2^4=16$ different sectors. However the sectors containing only one kind of Green's function (say, $G_{\uparrow n, \uparrow n}$ only) are empty. There is also a trivial sector corresponding to the zeroth order in the series. Consequently, in our calculation we performed a separate sampling other 11 non-trivial sectors, estimated their relative contributions, and obtained the final result as a weighted sum over all sectors. This allowed a significant extension of the temperature range available, although we still experience certain problems while lowering the temperature (see the next section).

\vskip0.5cm
\begin{section}
{\bf Results and discussion}
\end{section}
 We start from  discussing the results of  the calculations of pure neutron matter to check reliability of the  approximations made. It includes mapping of a  continuous system to the lattice  and applying the DMFT approach to the resulting Habbard model. Also, our interaction term is of a fairly simple structure and contains the $s$- wave only.

As a first check we have performed the calculations for the unitary Fermi gas (UFG) \cite{Gan,UFG}. UFG is characterized by infinite s-wave scattering length $a_s = \infty$ and zero effective radius $r_e = 0$.  Strictly speaking, neutron  matter can't be considered as being identical to UFG at any density, but at relatively low densities they are close and should display qualitatively the same behavior. 
Another reason to compare with UFG is that this system is very well studied both experimentally and theoretically. It has become possible to obtain UFG experimentally with the help of a Feshbach resonance \cite{Fesh}. This fact stimulated interest to this topic and quite accurate results were obtained for its properties using both theoretical and experimental methods \cite{Pit}.
%The parameter to be determined is $k_{max}$. Performing a benchmark %calculations for the system near the unitary limit (not shown in the %paper) we found that the best results are obtained for the lattice with %the filling factor of about $0.1$, which corresponds to $k_F\approx 0.4.. %0.5 k_{max}$. In this regime, the results are not sensitive to a %particular choice of $k_{max}$.

The central quantity to calculate in the unitary limit is the Bertsch parameter $\xi$ \cite{Ber} defining energy of the system as $E_{UFG} = \xi E_{FFG}$, where $E_{FFG} = \frac{3}{5} E_{F}$ is the energy of a free Fermi gas. Nowadays, very accurate results are available for both this parameter and for its deviation with small effective interaction range $k_F r_e \ll 1$, \cite{Car}. The corresponding analytic expression takes the form 
\begin{equation}
\label{UFGenergy}
\xi = 0.37 + 0.12 k_F r_e 
\end{equation}
To compare our system with the UFG we tune Hubbard interactions to infinite scattering length. 
Our  DMFT result is $\xi_{\text{DMFT}} = 0.44$ to be compared with the lattice result for real UFG with zero interaction range $\xi_{\text{UFG}} = 0.37$ \cite{Win}. The difference is $\frac{\xi_{\text{DMFT}} - \xi_{\text{UFG}}}{\xi_{\text{UFG}}} \approx 15\%$. That seems to be fairly reasonable given a number of the above mentioned assumptions. One notes, however that the lattice results also vary within 5-10 $\%$. For example, the fixed node Monte-Carlo simulation \cite{Ast} resulted in $\xi_{\text{UFG}} = 0.42$, much closer to the value obtained in this paper. The other observable we have looked at is the energy gap in unitary limit. Similar to energy density it can be expressed as $\Delta = \epsilon E_F$ where the constant $\epsilon$ is universal and does not depend on density. Our results for the gap are shown in Fig.1.

\begin{figure}[t!]
\includegraphics[width=0.75\linewidth]{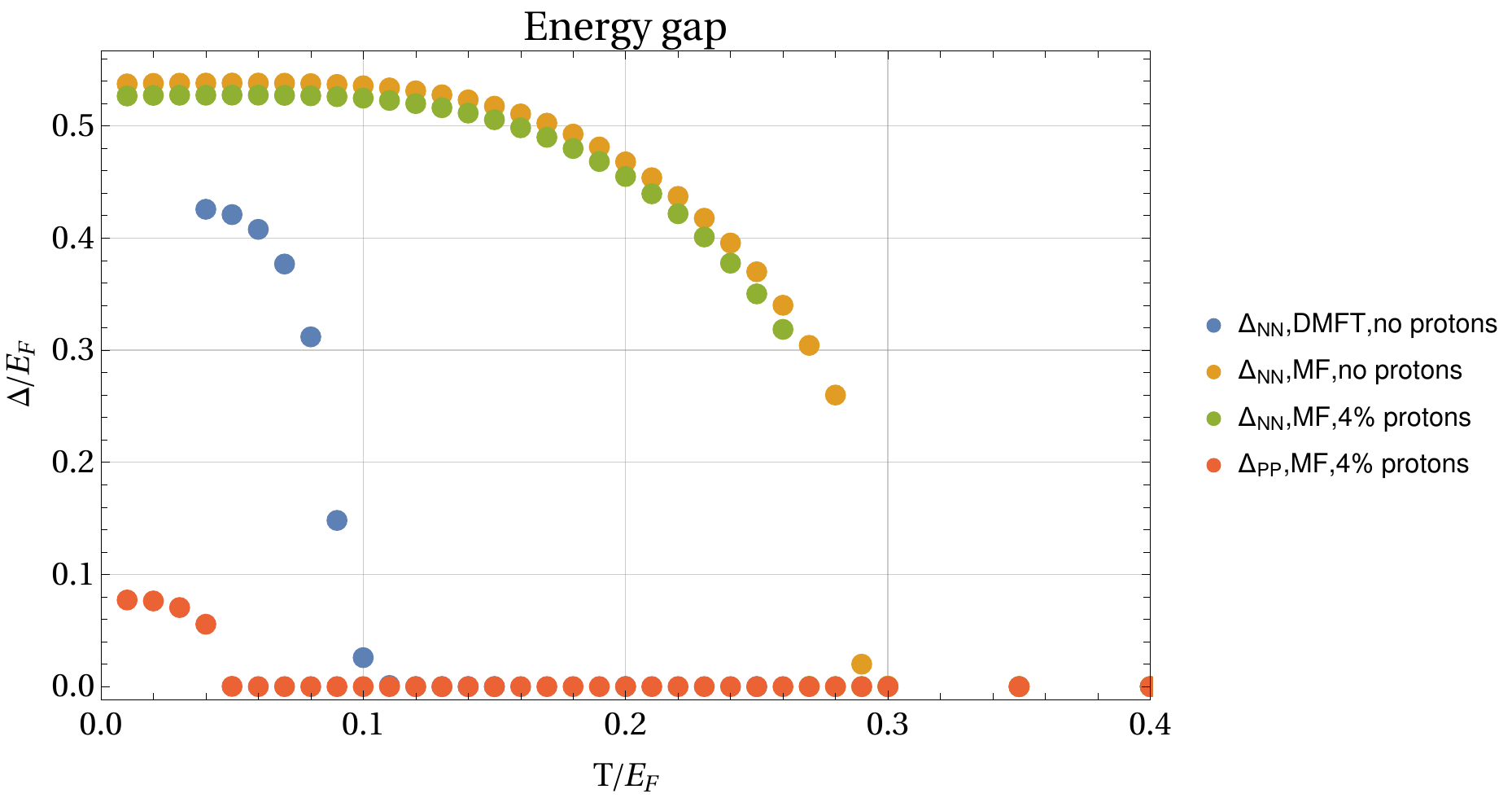} 
\caption{Energy gap in unitary limit with and without the proton fraction.}
\end{figure}
The calculations have been performed within both the MF and DMFT approaches with and without the proton fraction. For pure neutron matter the MF and DMFT calculations give $\epsilon =$ 0.86 and 0.7 correspondingly. Both values are comparable with those obtained in the other calculations \cite{Ugap} but a noticeable difference between them indicates the important role played by the correlations. Adding a small fraction of the protons does not change much the results for the MF calculations. Unfortunately we were not able to calculate the gaps within the DMFT approach when the proton fraction is included because of the large statistical errors at low temperatures where the corresponding gaps are supposed to be formed. As will be shown below the effect of the correlations becomes even more pronounces for the proton fraction related observables. One also notes that DMFT gives quite reasonable value of the critical temperature of the superfluid phase transition for the pure neutron matter, comparable with the other results \cite{Gan}. On the other hand  the MF approach provides the value of the critical temperature which seems too large to be realistic.

Now we turn to the results obtained for a realistic neutron matter with a finite scattering length and nonzero effective radius. We expect DMFT to perform even better in this case compared to the unitary limit as introducing lattice implies nonzero effective radius thus bringing DMFT closer to a real physical situation. The results obtained for the energy density are shown in Fig.2 together with the QMC calculations using AV4 nucleon-nucleon potential \cite {Arg}. As one can see from Fig.2 our DMFT results are quite close to the QMC ones thus demonstrating a significant improvement as compared to the MF results up to the densities $k_F \gtrsim 0.5 \text{fm}^{-1}$. At higher densities our description becomes progressively less and less reliable and at nuclear saturation density all we can hope is just a qualitative description of neutron matter. The physical reason is that our interaction Hamiltonian includes only the $s$ - wave part of the nucleon-nucleon forces whereas the contribution from higher partial waves becomes more and more important with increasing density.
\begin{figure}[t!]
\includegraphics[width=0.75\linewidth]{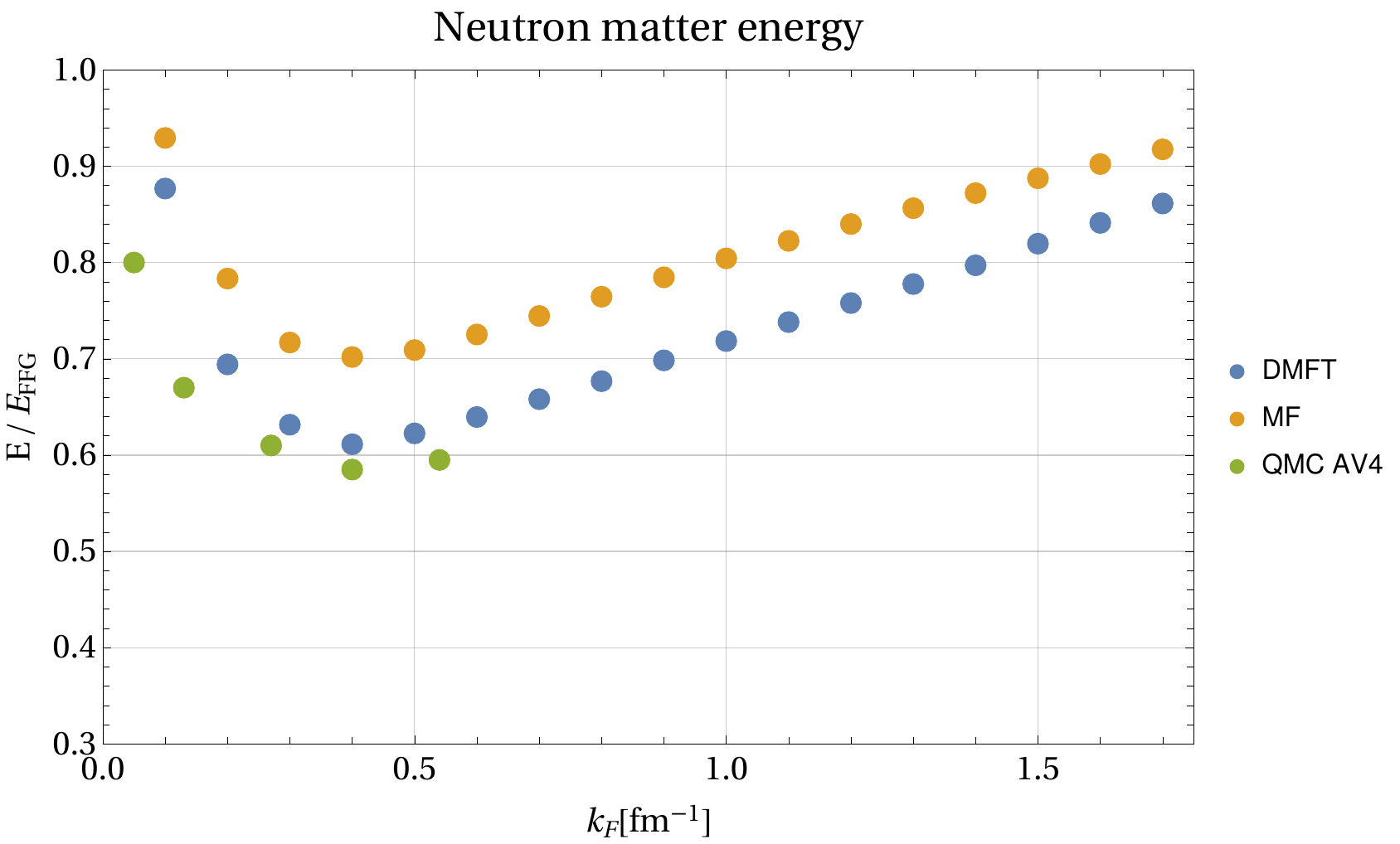} 
\caption{Energy density of realistic neutron matter. The blue and orange points correspond to the DMFT and MF calculations. The green points are the results of the QMC calculations with AV4 version of nucleon-nucleon potential.}
\end{figure}
\begin{figure}[t!]
\includegraphics[width=0.75\linewidth]{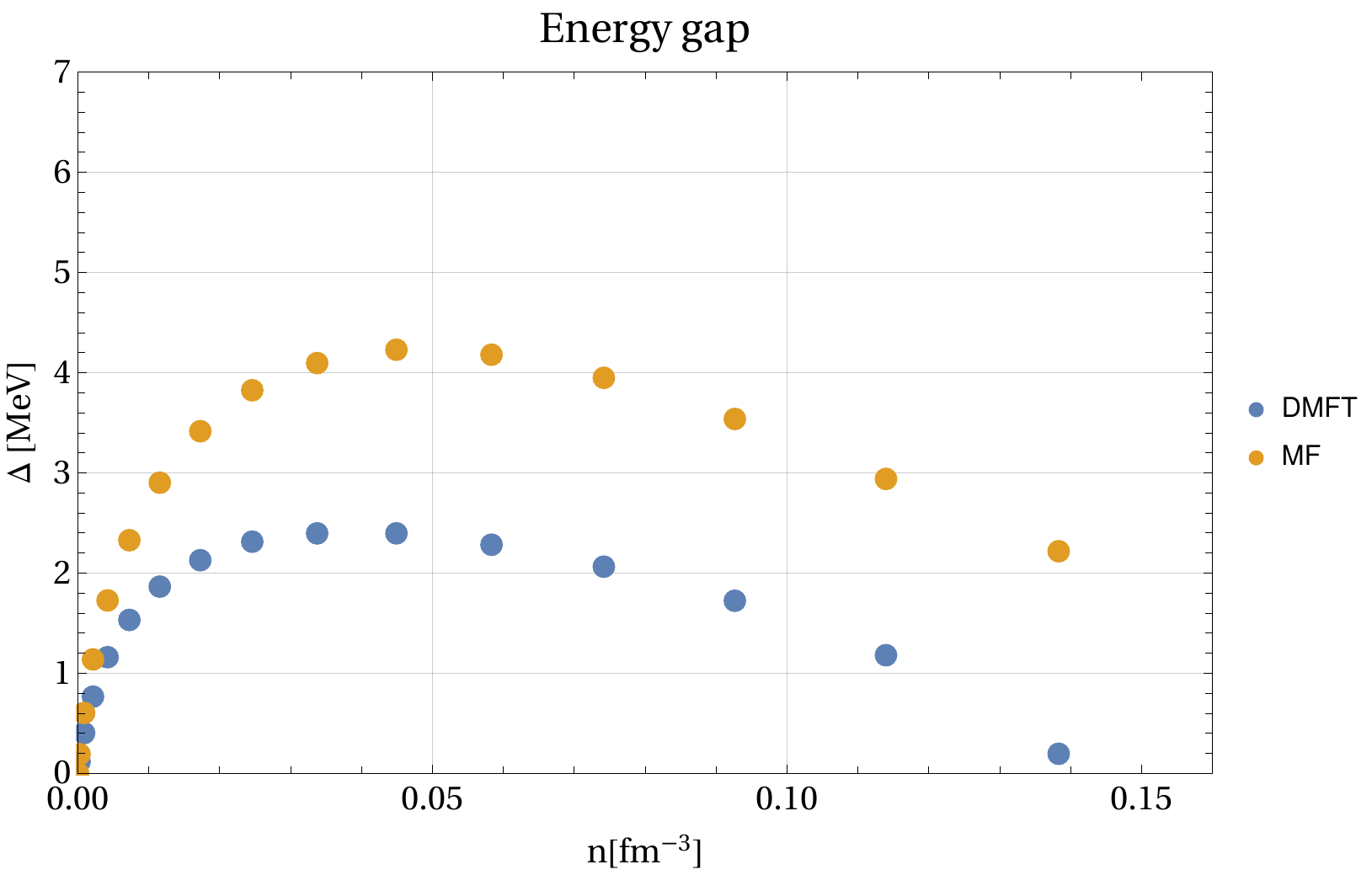} 
\caption{Energy gap of realistic neutron matter. The blue and orange points are DMFT and DM calculations correspondingly.}
\end{figure}
As an another check of a reliability of our approach we calculated the pairing gap of pure neutron matter. The results are shown in Fig.3. Again one can see that the DMFT calculations (blue dots online) gives the results for the gap much closer to those commonly accepted in the literature ($\Delta \simeq$ 1-2 MeV. On the contrary the MF approach (orange dots online) results in too large values of the gap.

After performing various validity checks we turn to asymmetric nuclear matter.
Before performing calculations for realistic neutron-proton matter, we again consider the unitary limit with interactions tuned to infinite scattering lengths. By doing this we again check a quality of the procedure which will be used for calculations of an asymmetric nuclear matter, namely an extrapolation  from low  to zero temperature. The reason is that  the lowest temperatures we are able to reach is $T = 0.15 E_F$ because of the statistical errors in Monte Carlo impurity solver so that some sort of extrapolation is required. In Figs.4,5  we show a dependence of energy on temperature for the MF and DMFT approaches, and for both pure neutron matter and neutron-proton matter with $4\%$ of protons. Only for DMFT calculations with protons we are limited to $T = 0.15 E_F$. In the DMFT calculations without protons it is possible to go to effectively zero energy. For all calculations except DMFT with protons we observe a linear behavior at high temperatures, and then sudden drop with a plateau afterwards. This drop is connected to appearance of the superfluid energy gap and the plateau is connected to the exponential behavior of energy at low temperatures due to a presence of the energy gap for quasi-particles $E(T) = E(0) + c_1 e^{-c_2 \frac{\Delta}{T}}$. Another important feature is that effect of adding protons on the energy is significantly bigger for the DMFT results, than it is for the MF ones implying a significant role played by the correlations in this case.   
\begin{figure}[t!]
\includegraphics[width=0.65\linewidth]{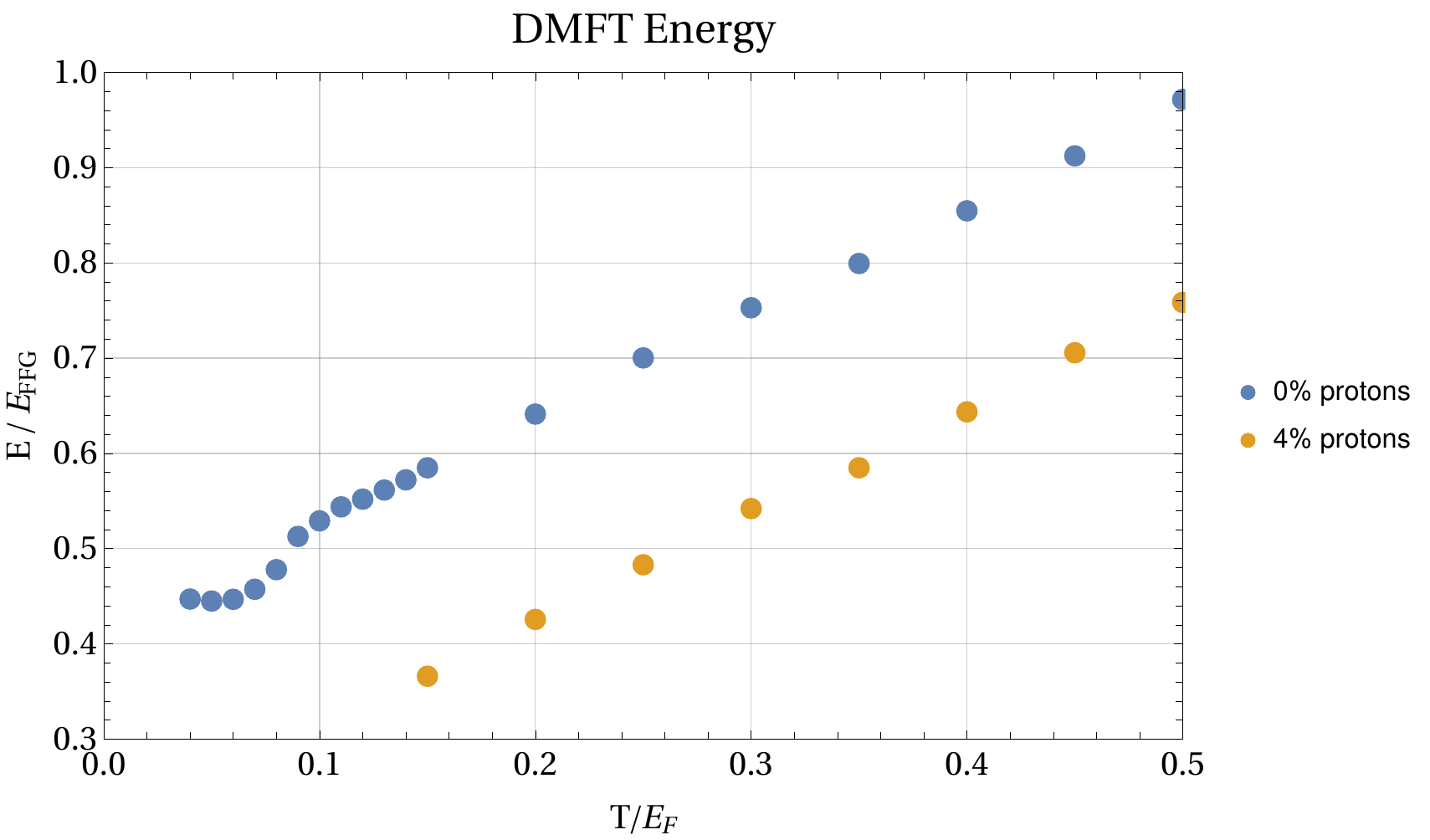} 
\caption{Energy density of asymmetric nuclear matter in DMFT approximation.}
\end{figure}
\begin{figure}[t!]
\includegraphics[width=0.65\linewidth]{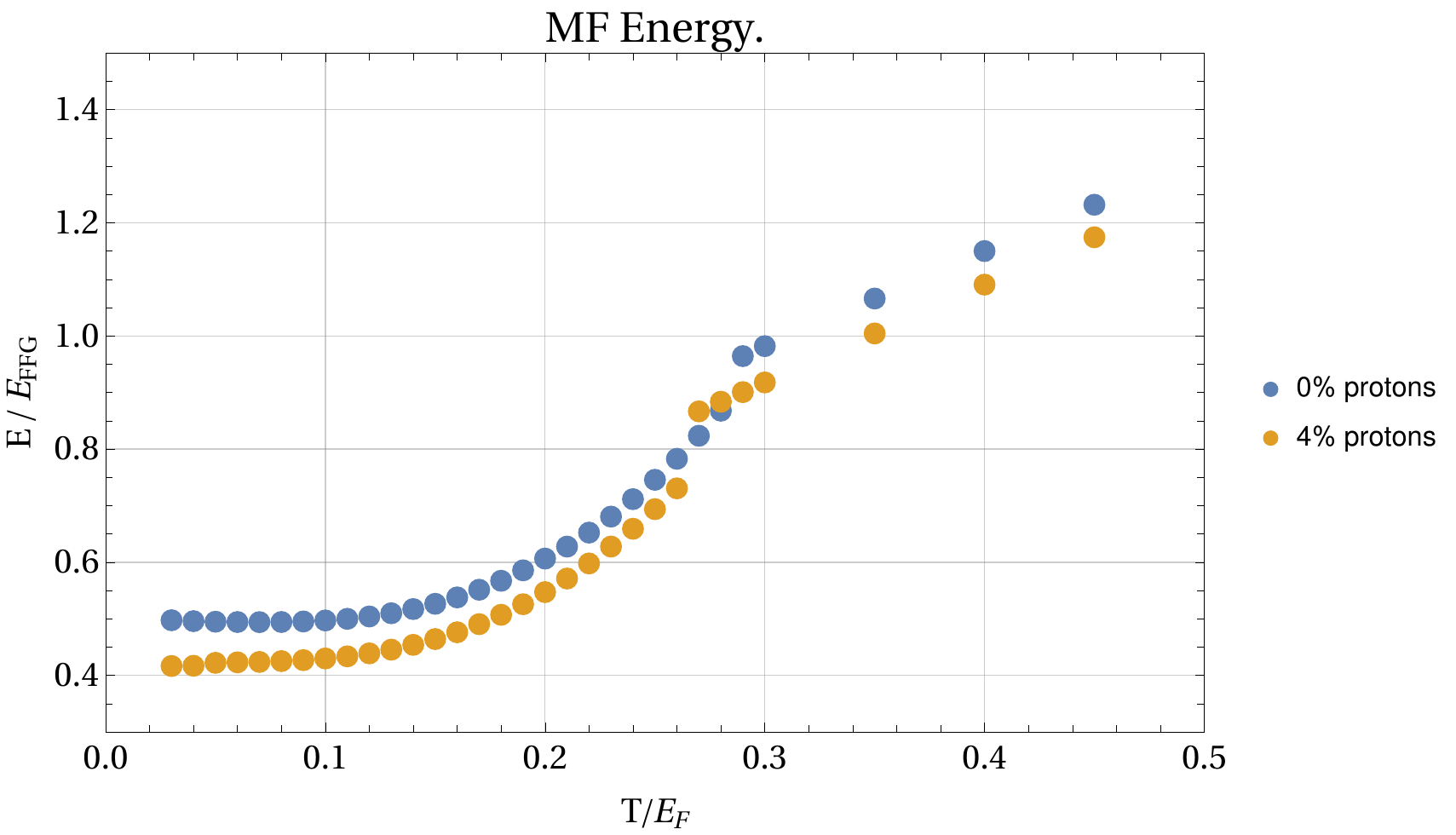} 
\caption{Energy density of asymmetric nuclear matter in MF approximation.}
\end{figure}

Now we turn to pairing. In principle we can have the pairings described by full $4\times4$ matrix $\Delta$. In our DMFT calculations we use the Nambu representation, where only the singlet neutron-proton pairing is taken into account.
\begin{figure}[t!]
\includegraphics[width=0.75\linewidth]{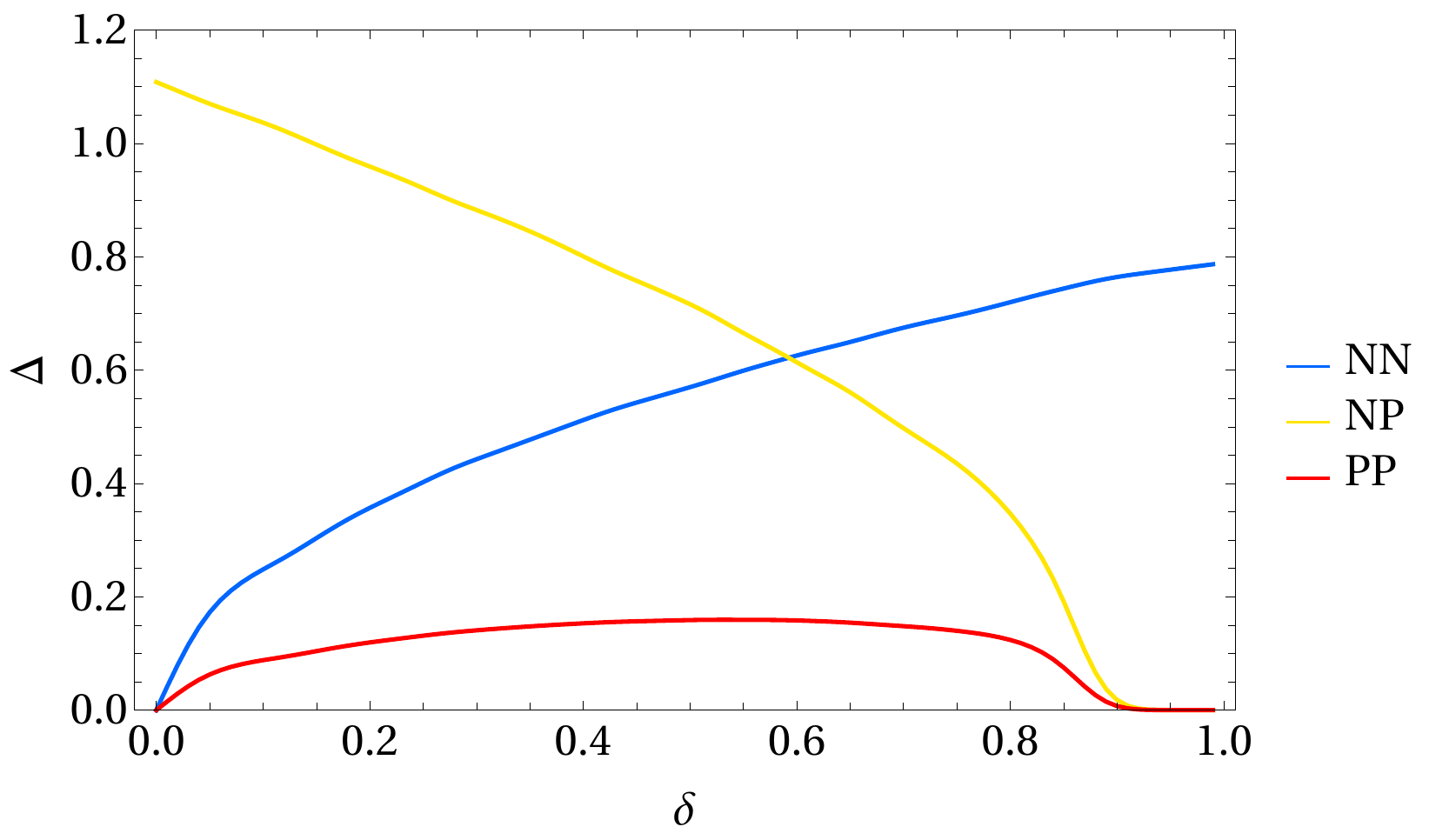} 
\caption{Energy gaps in neutron-neutron, neutron-proton and proton-proton channels. The asymmetry parameter $\delta$ is defined in the text}
\end{figure}

Including the triplet pairing in the DMFT is possible in principle but it is extremely hard to develop a reliable Monte Carlo solver. Therefore, the presence of the triplet neutron-proton pairing can be checked on MF level to validate neglecting it in the DMFT calculations. For density ranges from $\rho = 0.005 \rho_0$ to $\rho = 1.0 \rho_0$ and for all proton fractions we found out that self consistency MF equations for $\Delta$ converge to results with no neutron-proton triplet pairing.

In Fig.6 we show the results for all three paring gaps at the density $\rho = 0.01 \rho_0$ and full asymmetry range, which is defined as $\delta = \frac{n_n-n_p}{n_n+n_p}$. In the symmetric case $\delta = 0$ a neutron-proton gap dominates, but as we go to pure neutron matter with $\delta = 1$ a neutron-neutron gap becomes the only one. At all asymmetries, except close to $1$, proton-proton gap is present but it has relatively small magnitude. Also neutron-proton gap disappears at some critical value of asymmetry with this value decreasing with density.  This is well known effect stemming from the fact that the difference between proton and neutron Fermi spheres becomes too large to accommodate a standard zero-momentum pairing. Clearly, there is a possibility of formation of the neutron-proton pairs with a nonzero total momentum (LOFF pairing) but we do not consider such an option in this paper. 

In Fig.7 we show the results of the DMFT calculations  for the energy density of an asymmetric matter as a function of density at the different values of the proton fraction. One can see that although adding protons lowers the energy the overall tendency remains the same for all proton fractions and the curves differ by just some scaling factor. Our DMFT calculations were performed at temperature $T = 0.3E_F$ and for range of proton fraction from 2$\%$ up to  10$\%$   
\begin{figure}[t!]
\includegraphics[width=0.75\linewidth]{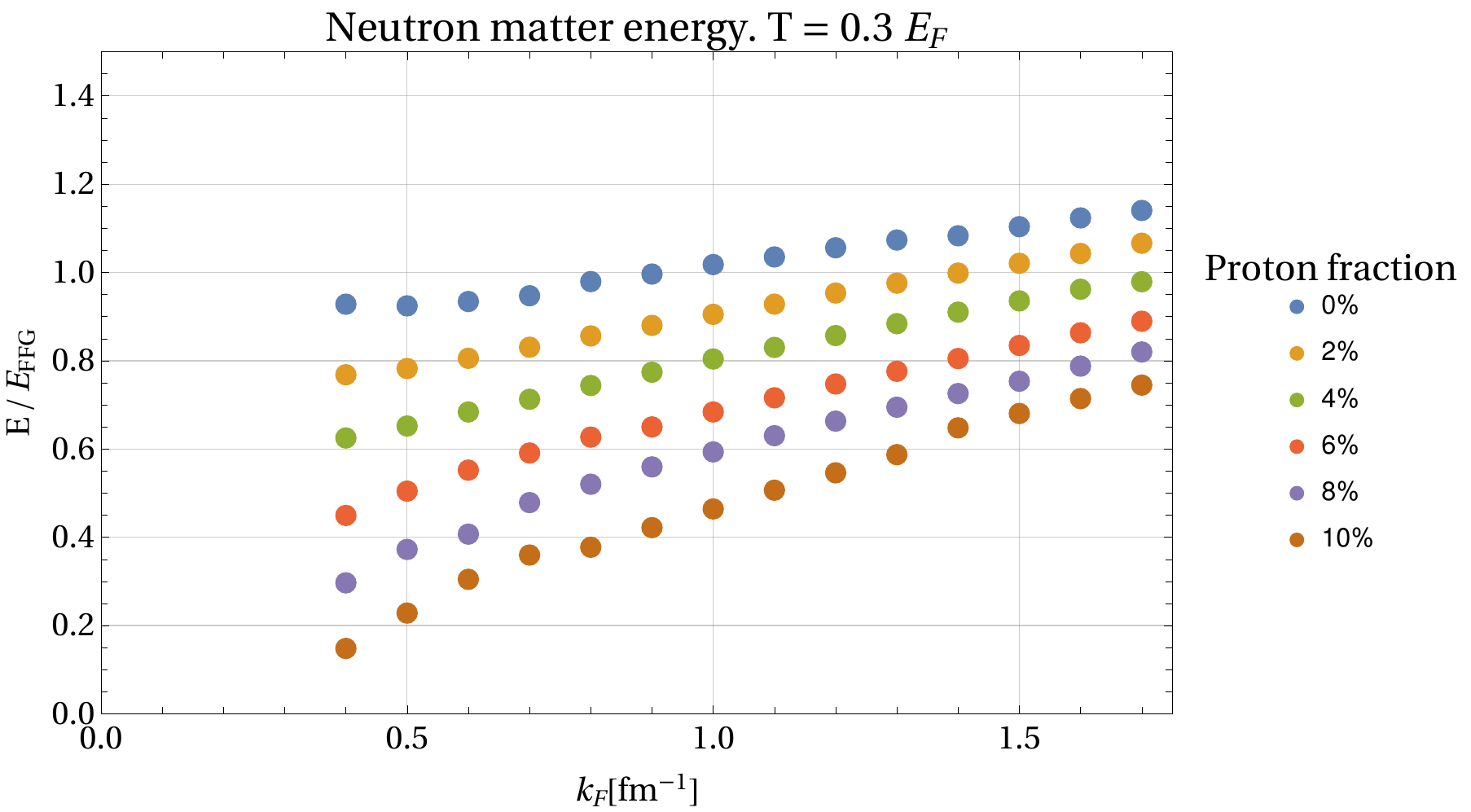} 
\caption{Energy density of asymmetric nuclear matter. The asymmetry parameter $\delta$ is defined in the text}
\end{figure}

\vskip1cm
\begin{section}
{\bf NUCLEAR MATTER IN BETA-EQUILIBRIUM}
\end{section}

Here we will consider neutron matter at the density range corresponding to Fermi momentum $0.4 \; \text{fm}^{-1} \lesssim k_F \lesssim 1.7 \; \text{fm}^{-1}$. However, as explained above  the density range where our results are the most reliable is $0.5 \; \text{fm}^{-1} \lesssim k_F \lesssim 1.0 \; \text{fm}^{-1}$ which roughtly corresponds to the densities in the inner crust/outer core of a neutron star. The matter in the star is in beta-equilibrium so that proton fraction has to satisfy this condition.  It can be estimated from the analytic expression where the independence of the chemical potential from the proton fraction is assumed

\begin{equation}
\nu \approx \left(\frac{k_F}{9.54} \frac{\mu_n - \mu_p}{\mu_{FG}}\right)^3
\end{equation}

 The corresponding results for both DMFT and MF approaches are shown on Fig.8. One can see that that taking into account correlations changes the picture significantly so that  the proton fraction obtained in the DMFT approach is  larger than that in the MF one. Whereas the MF approach gives a vanishingly small proton fraction, especially at low densities, the proton fraction in DMFT is approximately $1\%$ for $k = 0.5 \; \text{fm}^{-1}$ and reaches $\simeq 8\%$ at $k_F = 1.7 \; \text{fm}^{-1}$. Again, this results indicate that taking into account correlations is extremely important especially for the proton observables. Interestingly enough, the proton fraction at saturation density is quite close the commonly accepted value of approximately $5\%$ and grows quite fast with increasing density reaching at $k_F = 1.7 \; \text{fm}^{-1}$ the value  which is not far from one at which the direct URCA process begins to operate. Clearly, given insufficient reliability of our approach at normal nuclear density and above the exact value of an asymmetry can't be interpreted as a quantitative estimate but the conclusion about the importance of the correlations in determing the proton fraction seems to be quite robust. To get a reliable estimate of this quantity at higher density one need to use more sophisticated treatment which would include taking into account the effect of higher partial waves in the nucleon-nucleon interaction as well as three-body forces which play increasingly important role at larger densities. However, our results indicate that URCA process can start at lower density than it is usually assumed. This is an interesting observation which clearly deserves further studies.

\begin{figure}[t!]
\includegraphics[width=0.75\linewidth]{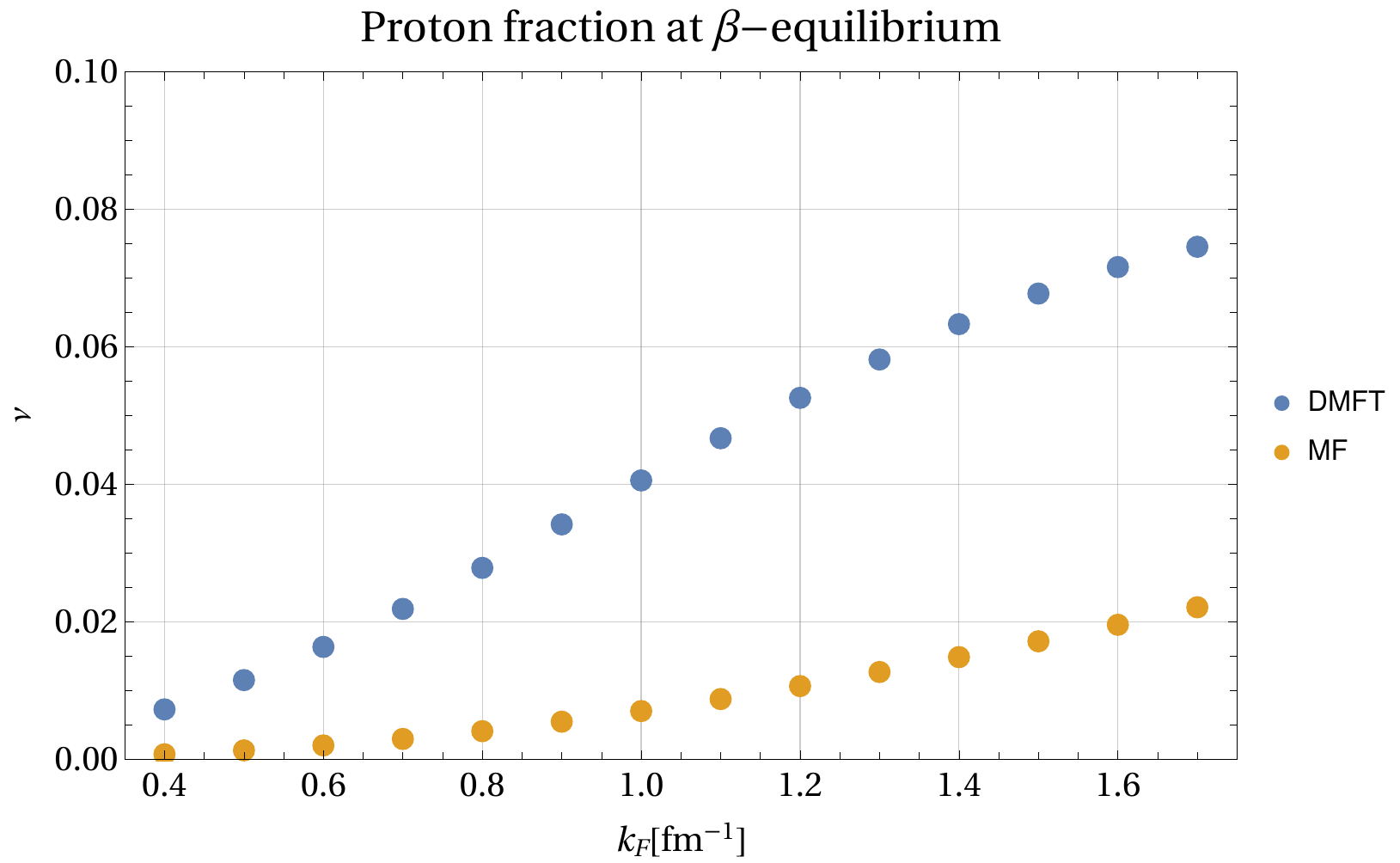} 
\caption{Proton fraction in beta-equilibrium.}
\end{figure}
%\begin{figure}[H]
%	\includegraphics[width=0.8\textwidth]{images/5-7-7.pdf}
%	\caption[Neutron matter energy for different proton fractions]{Neutron matter energy for different proton fractions. This is calculations are performed at nonzero temperatures and later extrapolated to $0$ temperature}
%	\label{fig:DMFTbareresults}
%\end{figure}
\begin{figure}[t!]
\includegraphics[width=0.75\linewidth]{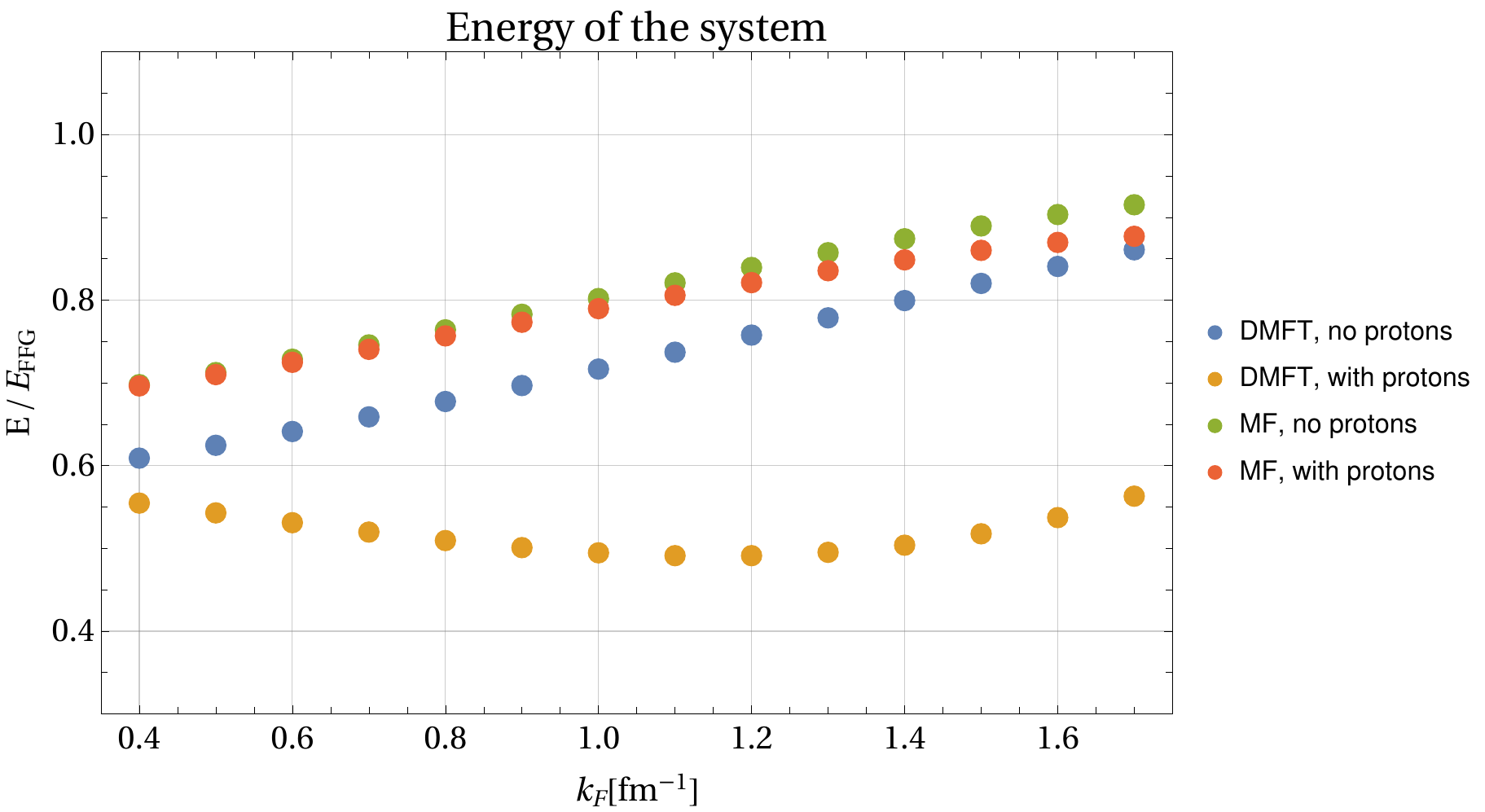} 
\caption{Energy of neutron matter in beta-equilibrium.}
\end{figure}

Using the results for equilibrium proton fraction  we calculate the energy of neutron matter in beta-equilibrium as well as the pressure. The pressure is calculated as a derivative of energy with respect to density. The results for the energy density calculated both in the DMFT and MF approximation are shown in Fig.9.
%\begin{figure}[H]
%	\includegraphics[width=0.5\textwidth]{images/5-7-4.pdf}
%	\includegraphics[width=0.5\textwidth]{images/5-7-5.pdf}
%	\caption[Energy and pressure of realistic neutron matter in beta-equilibrium]{Energy and pressure of realistic neutron matter in beta-equilibrium}
%	\label{fig:DMFT-MFenergypressure}
%\end{figure}
 
\begin{figure}[t!]
\includegraphics[width=0.7\linewidth]{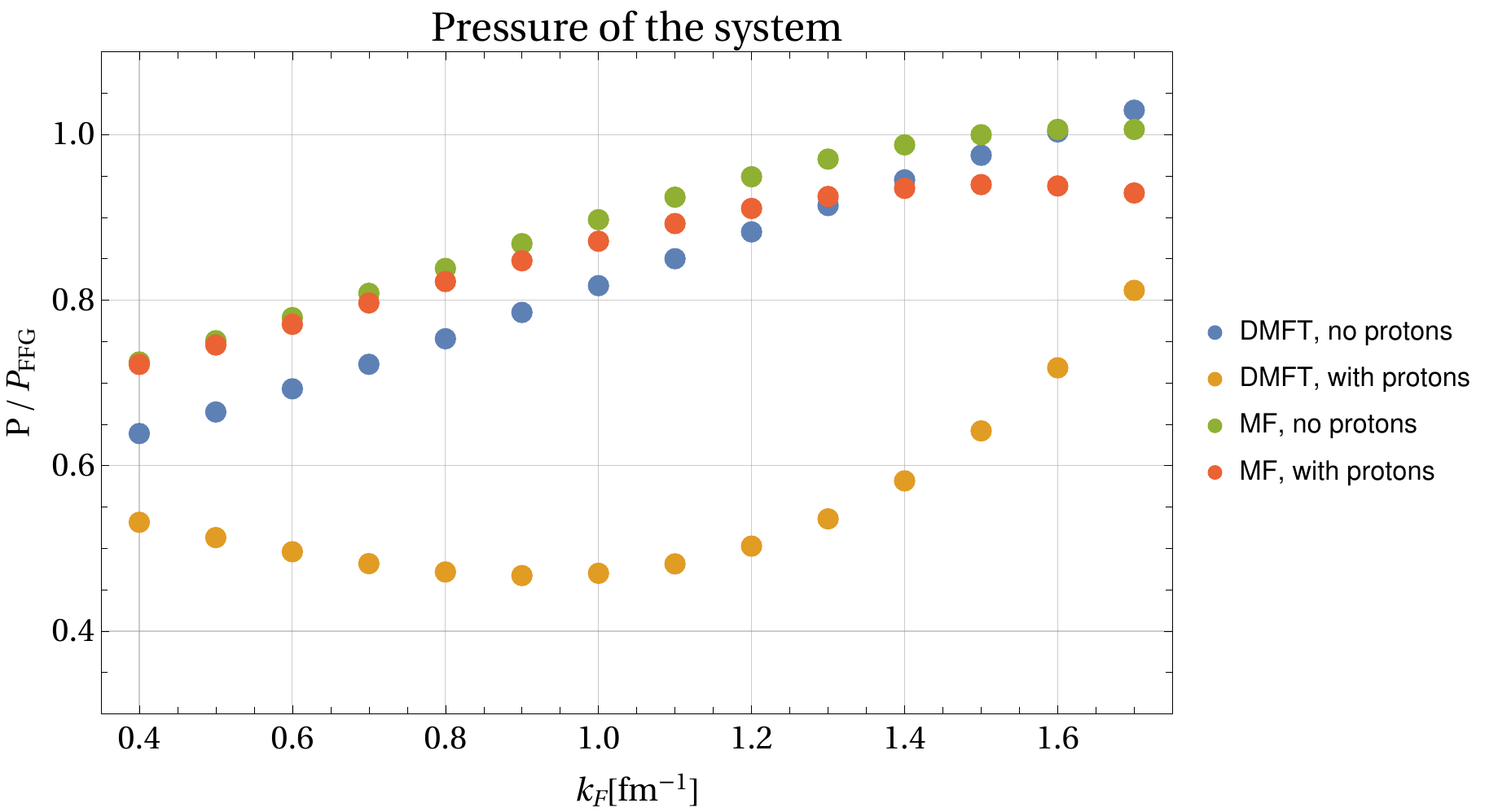} 
\caption{Pressure of neutron matter in beta-equilibrium.}
\end{figure}

\newpage
 We have already seen that taking into account the local correlations in DMFT significantly changes the quantities related to the proton fractions. The contribution of protons to the energy density of the system in beta-equilibrium is a product of equilibrium proton fraction and its energy, therefore the effect of correlations is in some sense amplified for this quantity. This results in a large corrections when the  proton contribution to energy density is taken into account. Similar conclusion holds for the pressure shown in Fig.10 where the effect of correlations is pronounced even more.

 Moreover, for the densities with $k_F \gtrsim 0.6 \; \text{fm}^{-1}$ the contribution of protons to energy density becomes larger than the contribution of correlations to the energy density of pure neutron matter. At this point it appears desirable to include corrections which are related to the nonlocal (momentum dependent) part of self-energy. This is doable in principle, however it is technically challenging, so is relegated to future studies.

\vskip0.5cm
\begin{section}
{\bf CONCLUSION}
\end{section}
     Lattice model of a asymmetric nuclear matter is constructed.  Our lattice Hamiltonian  is just a Hubbard model with a local on-site interaction. We have tuned the parameters of nucleon-nucleon interaction  using the experimental data on the s-wave scattering phase shifts at the specially chosen momentum inside the Fermi surface. The DMFT approach has been used to calculate the observables and study a role of the correlations. 
\par
 To solve impurity problem inside Dynamical Mean Field theory we use the Quantum Monte Carlo solver approximation. 
To validate our approach we performed calculations for pure neutron matter, for which accurate results are available. The calculations have been done for both  the unitary limit and realistic neutron matter. From this comparison we determined that DMFT captures most of the correlations in the system and quite accurately calculates such characteristics as energy density and the Bertsch parameter. Most errors of the whole scheme come from replacing originally continuous neutron matter with Hubbard model. However in the density region $0.025\rho_0 \lesssim \rho \lesssim 0.2 \rho_0$ the results are quite reliable. By performing both the DMFT and MF calculations we were able to estimate the effect of correlations since the former takes this into account whereas the latter does not. For pure neutron matter properties the effect of correlations was moderate, responsible for $10-30\%$ corrections to the observables. It in general agrees with the findings from \cite{Pri} where the DMFT approach has been applied to the Fermi gas at unitarity. For the proton related quantities - the proton fraction at equilibrium and the proton energy the effect of correlations was found to be much more important, modifying significantly the MF results. In particular, we have found large changes in the density dependencies of the energy and pressure of the neutron matter in beta-equilibrium when a presence of the proton impurity is taken into account. The proton fraction itself has been determined using the beta-equilibrium requirement. Our DMFT results taking into account the correlation effects turned out to give the value of the proton fraction still below but much  closer to the critical one, required to initiate the direct URCA process. Overall one can conclude that taking into account a proton fraction together with the correlation effects  are essential for a realistic description of neutron stars at the transitional region between the inner crust and outer core. 
\par
There are several ways for improving and further developing our approach. First, one would like to improve our Monte Carlo code to reduce statistical fluctuations. Second, the three-body forces should be included. It is rather well established that the three-body forces becomes progressively more and more important when increasing density. Third it would be interesting to include higher partial waves in our model for nucleon-nucleon interactions as well as the contribution of the nonlocal correlations using for example some recently developed generalisations of DMFT \cite{Rub} or functional renormalisation group \cite{Kri}
The progress along these lines will be reported in future publications.

\end{document}